\documentstyle[12pt,epsf]{article}

\newfont{\lfont}{line10}

\newcommand{\BE}{\begin{equation}}
\newcommand{\EE}{\end{equation}}
\newcommand{\BEA}{\begin{eqnarray}}
\newcommand{\EEA}{\end{eqnarray}}

\def\p{\partial}

\def\12{\frac{1}{2}}
\def\bea{\begin{eqnarray}}
\def\eea{\end{eqnarray}}
\def\ba{\begin{array}}
\def\ea{\end{array}}

\def\one-loop{\mbox{\scriptsize one-loop}}

\def\a{\alpha}

\def\G{\Gamma}

\catcode`\@=11

\def\@normalsize{\@setsize\normalsize{15pt}\xiipt\@xiipt
\abovedisplayskip 14pt plus3pt minus3pt%
\belowdisplayskip \abovedisplayskip
\abovedisplayshortskip  \z@ plus3pt%
\belowdisplayshortskip  7pt plus3.5pt minus0pt}

\def\small{\@setsize\small{13.6pt}\xipt\@xipt
\abovedisplayskip 13pt plus3pt minus3pt%
\belowdisplayskip \abovedisplayskip
\abovedisplayshortskip  \z@ plus3pt%
\belowdisplayshortskip  7pt plus3.5pt minus0pt
\def\@listi{\parsep 4.5pt plus 2pt minus 1pt
            \itemsep \parsep
            \topsep 9pt plus 3pt minus 3pt}}

\def\underline#1{\relax\ifmmode\@@underline#1\else
        $\@@underline{\hbox{#1}}$\relax\fi}
\@twosidetrue

\relax

\catcode`@=12

\catcode`\@=11

\def\section{\@startsection{section}{1}{\z@}{3.5ex plus 1ex minus
   .2ex}{2.3ex plus .2ex}{\large\bf}}

\def\thesection{\Roman{section}.}

\def\appendix{\setcounter{section}{0}
        \def\thesection{Appendix }
        \def\theequation{\Alph{section}.\arabic{equation}}}

\def\ps@headings{\def\@oddfoot{}\def\@evenfoot{}
\def\@oddhead{\hbox{}\hfill
        \makebox[.5\textwidth]{\raggedright\ignorespaces --\thepage{}--
        \hfill {}}}
\def\@oddhead{\hbox{}\hfill --\thepage{}-- \hfill
        {}}
\def\@evenhead{\@oddhead}
\def\subsectionmark##1{\markboth{##1}{}}
}

\ps@headings

\catcode`\@=12

\relax

\def\figcap{\section*{Figure Captions\markboth
        {FIGURECAPTIONS}{FIGURECAPTIONS}}\list
        {Fig. \arabic{enumi}:\hfill}{\settowidth\labelwidth{Fig. 999:}
        \leftmargin\labelwidth
        \advance\leftmargin\labelsep\usecounter{enumi}}}
 \relax
\def\tablecap{\section*{Table Captions\markboth
        {TABLECAPTIONS}{TABLECAPTIONS}}\list
        {Table \arabic{enumi}:\hfill}{\settowidth\labelwidth{Table 999:}
        \leftmargin\labelwidth
        \advance\leftmargin\labelsep\usecounter{enumi}}}
 \relax
\def\reflist{\section*{References\markboth
        {REFLIST}{REFLIST}}\list
        {[\arabic{enumi}]\hfill}{\settowidth\labelwidth{[999]}
        \leftmargin\labelwidth
        \advance\leftmargin\labelsep\usecounter{enumi}}}
 \relax

\catcode`\@=11

\def\ps@headings{\def\@oddfoot{}\def\@evenfoot{}
\def\@oddhead{\hbox{}\hfill
        \makebox[.5\textwidth]{\raggedright\ignorespaces --\thepage{}--
        \hfill {}}}
\def\@evenhead{\@oddhead}
\def\subsectionmark##1{\markboth{##1}{}}
}

\ps@headings

\relax

\newskip\humongous \humongous=0pt plus 1000pt minus 1000pt

\newif\ifdtup

\def\beq{\begin{equation}}
\def\eeq{\end{equation}}

\def\beqn{\begin{eqnarray}}
\def\eeqn{\end{eqnarray}}
\relax

\def\G2{{\; \rm GeV/}c2}
\def\G{\; \rm GeV}

\def\dotx{\dotx{\dot\overline{x}}}

\relax

\def\Bar{\overline}
\def\p{\partial}

\begin{document}
%
%
\begin{titlepage}

\renewcommand{\thefootnote}{\fnsymbol{footnote}}

\begin{flushright}
        \normalsize
      April, 2003 \\
    OCU-PHYS 198 \\
    hep-th/0304184  \\

\end{flushright}

%
\begin{center}
     {\large\bf  Supereigenvalue Model  and \\
        Dijkgraaf-Vafa  Proposal}
\footnote{This work is supported in part
   by the Grant-in-Aid  for Scientific Research
(14540264, 14540073) from the Ministry of Education,
Science and Culture, Japan.}
\end{center}

\vfill

\begin{center}
      {%
H. Itoyama\footnote{e-mail: itoyama@sci.osaka-cu.ac.jp}
\quad and \quad
H. Kanno\footnote{e-mail: kanno@math.nagoya-u.ac.jp}
}\\
\end{center}

\vfill

\begin{center}
        ${}^{\dag}$\it  Department of Mathematics and Physics,
          Graduate School of Science\\
          Osaka City University\\
          3-3-138, Sugimoto, Sumiyoshi-ku, Osaka, 558-8585, Japan  \\
\vskip 1.0em
        ${}^{\ddag}$\it Graduate School of Mathematics, Nagoya University \\
         Chikusa-ku, Nagoya, 464-8602, Japan  \\
\end{center}

\vfill

\begin{abstract}

   We present a variant of the supereigenvalue model proposed before
   by Alvarez-Gaume, Itoyama, Manes, and Zadra. This model derives a set of
  three planar loop equations which takes the same form as the set
 of three anomalous  Ward-Takahashi identities on the gaugino
  condensates recently derived by Cachazo, Douglas,
   Seiberg and Witten. 
    Another  model which implements ${\cal N}=2$ superVirasoro constraints
  is constructed for comparison.

\end{abstract}

\vfill

\setcounter{footnote}{0}
\renewcommand{\thefootnote}{\arabic{footnote}}

\end{titlepage}

Over the decades matrix models have contributed to our understanding
  of gauge fields and strings and their synthesis in different contexts.
 The recent proposal of Dijkgraaf and Vafa \cite{DV},
  (see \cite{prehistory} for prehistory to this),
  has enabled us to relate exact results on the superpotential
 and gauge couplings
  for a wide class of ${\cal N}=1$ supersymmetric gauge theories
 with the planar limit of a matrix model in which the
 superpotential is taken as an ordinary potential.
 One may regard this relation as that connecting
  two faces of the  notion of integrability of these theories in
 their infrared limit.
  The one face is seen as the integrability of
 Seiberg-Witten, ${\it i.e.}$, ${\cal N}=2$ theories \cite{SW}
\cite{SWINT}
with soft breaking superpotential as Whitham  flows
\cite{whitham}.  Here
the aspects associated with gaugino condensates
 are not revealed immediately.
The other face is, of course, the integrability of matrix models
 where the (planar) free energy provides an integral
 representation of
   the prepotential. Here, bosonic integrations  of matrices
  have no way to capture the properties associated with
 supersymmetry and 
  we have to input a given breaking pattern of the original
  gauge theory. 
It seems that  we need to regard both gaugino condensates
 and vacuum values as moduli in order to gain unified
 understanding of this subject \cite{IM4}. 
Studies of the Dijkgraaf-Vafa
 proposal from these perspectives have been continuing
 both on its structure  and 
explicit calculations \cite{Gopa, NSW,IM4, DVint}. For the other aspects
 on this subject, see an extensive list of references seen,
 for instance,  in \cite{IM4} and the last reference of \cite{DVint}.
 
The proposal of Dijkgraaf and Vafa
 has been put and consolidated in a field theoretic vein
  by Cachazo, Douglas, Seiberg and Witten \cite{CDSW}
 in the case of an adjoint chiral superfield and an
 arbitrary superpotential.
In particular,  they have derived 
 an equation of the anomalous
 Ward-Takahashi identity associated with the Konishi anomaly \cite{Konishi}
 which takes the same form as the loop equation of the ordinary
 hermitian one-matrix model in the planar limit. 
  In their discussion,
there is actually a system of three Ward-Takahashi identities 
 involving
generating functions $R(z), w_\a(z), T(z)$, which are  all
 chiral operators related by  constant fermionic shift of the 
gaugino field strengths.\footnote{
This fermionic shift is allowed by the decoupling of the $U(1)$ 
factor in the adjoint representation.}
The first operator $R(z)$ realizes the proposal of \cite{DV}
   while the third operator $T(z)$ is closely related to the Seiberg-Witten differential. \footnote{This latter statement holds even
 in those cases in which unbroken nonabelian gauge symmetry 
survive.}
The system of three Ward-Takahashi identities appears
 to offer a unified understanding of \cite{DV} and
 ${\cal N}=2$ Seiberg-Witten theory. See \cite{CSW} for a recent 
progress.
  In the ordinary bosonic one-matrix model, on the other hand, 
there is only one kind of resolvent whose Laplace transform is 
a loop  operator, and is  no hope within this model to obtain 
the second,
which is fermionic, and the third resolvents carrying lower dimensions
than the first one. A model which derives all of the
  three equations of \cite{CDSW} and which provides a
 constructive definition serving both for the planar processes and
 for the non-planar (and therefore) gravitational ones
 has remained elusive.

To incorporate the counterpart of the fermionic shift operation,
fermionic degrees of freedom are required.
  The supereigenvalue model proposed sometime ago in \cite{AIMZ}, which
   materializes discretized two-dimensional supergravity,
  (see \cite{AIMZfollowup} for some of the subsequent developments), 
has come to our attention.\footnote{Supermatrices have been known not to
 serve well in the old context of $2d$ supergravity. 
(See, for instance, \cite{supermat} for this).
 For discussions in the present context, see \cite{supermatrecent}. 
There is a subtle issue of a decoupling of the 
superunitary group $U(M|N)$
in the integrations over the supermatrix. These 
integrations in general do not reduce
to those of the (super)eigenvalues. 
Furthermore,
  supertrace  requires us to introduce
 oppositely charged particles
 in the Dyson gas description of the (super)eigenvalues.
The supermatrices appear to us more appropriate to 
the quiver gauge theories and not to
the pure Yang-Mills. }
   In this letter, we will present a variant of the supereigenvalue
   model in which  the ordinary grassmann ${\cal N}=2$ superspace
   and its supercoordinates
  $(\lambda, \theta, \bar{\theta})$ \footnote{ $\theta$ and $\bar{\theta}$
   are two independent grassmann  variables and no complex conjugation 
is involved.}
  are introduced to label  a set of
   eigenvalues  $\lambda_{i}$ and their grassmann partners
$\theta_{i}, \bar{\theta}_{i}$. We will show  that this model
  captures the three equations of
  \cite{CDSW} as planar loop equations.

Let us first consider the following partition function:
\beqn
\label{Z}
  Z_{\check{N}} = \int \cdots \int \prod_{i=1}^{\check{N}} d 
\lambda_{i} d\theta_{i}
  d\bar{\theta}_{i}  \prod_{i < j} ( \lambda_{i} -\lambda_{j} +
  \theta_{i}\bar{\theta}_{i} -\theta_{j}\bar{\theta}_{j} )^{2}
  \exp \left( -\frac{1}{g_{s}}  \sum_{i}
  {\bf V} (\lambda_{i}, \theta_{i}, \bar{\theta}_{i} ) \right) \;\;,
\eeqn
   where   $g_{s} \equiv \frac{S}{\check{N}}$ and
\beqn
\label{bfV}
{\bf V} (\lambda_{i}, \theta_{i}, \bar{\theta}_{i} )
  =    W( \lambda_{i}; g_{\ell})
+  \Bar{W}^{(s)}( \lambda_{i}; \bar{\xi}_{\ell}) \theta_{i}
+   \bar{\theta}_{i} W^{(s)}( \lambda_{i}; \xi_{\ell})
+    \widetilde{W}( \lambda_{i}; \tilde{g}_{\ell})
  \theta_{i}\bar{\theta}_{i} \;\;
\eeqn
  is an ${\cal N}=2$ superfield in the
  terminology of the two-dimensional superconformal field theory
    in which the coefficients of the Taylor expansion in $\lambda_{i}$
  of the components $W, \Bar{W}^{(s)}, W^{(s)}, \widetilde{W}$
  are realized as
   the couplings $g_{\ell}, \bar{\xi}_{\ell}, \xi_{\ell}, \tilde{g}_{\ell}$.
  All of the singlet observables are obtained from this partition function
  by acting upon these couplings.  We have fixed the
  normalization  of the grassmann variable, so that the relative
   coefficient between
   $\lambda_{i} -\lambda_{j}$ and
  $\theta_{i}\bar{\theta}_{i} -\theta_{j}\bar{\theta}_{j} $
  in the measure factor is $1$. \footnote{
  In this letter, we  set the value of the exponent of the
  measure factor to $2$. For a general value $\beta$, the coefficient of
 an order $g_{s}$
  term in the loop equation below gets modified after a rescaling of
 the superpotential and the resolvent.}

  We bring this model into more direct contact to the proposal of
  \cite{DV}
  and  its description as gauge theory in \cite{CDSW}
   by reducing this partition function. Let us set
\beq
   \Bar{W}^{(s)}= W^{(s)}=0, \;\; \widetilde{W} =W^{\prime} \;\;,
\eeq
  which is, of course, a  nonsupersymmetric reduction.
  We regard the function $W$  an $n$-th order polynomial in its argument.
  The model obtained  this way is
\beqn
\label{ZIK}
  Z_{\check{N}} = \int \cdots \int \prod_{i=1}^{\check{N}} d 
\lambda_{i} d\theta_{i}
  d\bar{\theta}_{i}  \prod_{i < j} ( \lambda_{i} -\lambda_{j} +
  \theta_{i}\bar{\theta}_{i} -\theta_{j}\bar{\theta}_{j} )^{2}
  \exp \left( -\frac{1}{g_{s}}
  \sum_{i}  W (\lambda_{i} + \theta_{i}\bar{\theta}_{i}) \right) \;\;.
\eeqn

There is another supereigenvalue model which possesses ${\cal N}=2$ 
supersymmetry,
which we will present briefly in the end of this letter.
In contract with the model (\ref{ZIK}), the model with ${\cal N}=2$ 
supersymmetry
is obtained by a different choice of the  measure factor  
$\prod_{i < j} ( \lambda_{i} -\lambda_{j} -
 (\theta_{i}\bar{\theta}_{j} + \bar\theta_{i} \theta_{j}) )^{2}$.
\footnote{These two alternative choices may be characterized as
  follows: by requiring
the antisymmetry in $(i,j)$ and ${\cal N}=2$ neutralness, possible bilinear
combinations are either $\theta_{i}\bar\theta_{i} - 
\theta_{j}\bar\theta_{j}$ or
$\theta_{i}\bar{\theta}_{j} + \bar\theta_{i}\theta_{j}$. The 
difference is that
the former is odd under the exchange of $\theta$ and $\bar\theta$  while
the latter is even. Thus the former may be called chiral choice and the latter
non-chiral. In fact the combinations $\lambda \pm \theta\bar\theta$ are the
natural coordinates for chiral and anti-chiral superfields
 respectively.}

Let us briefly recall the logic which connects the Virasoro constraints
  at finite $\check N$ with its loop equation in the ordinary
  hermitian  one-matrix model\cite{Virasoroconst}.
  Consider the partition function with all grassmann degrees of
  freedom removed in eq.(\ref{Z}).   Setting up the system of
  Schwinger-Dyson equations associated with most general (polynomial)
variations of the matrix is equivalent to inserting the
   Virasoro operators (which is a total derivative)
   $\ell_{m}^{(B)} \equiv  -
  {\displaystyle \sum_{i=1}^{\check{N}} }
\frac{d}{d\lambda_{i}} \lambda_{i}^{m+1} \;,  \;\; m \geq -1$
  in the partition function and equating the resulting expressions
   to  zero. Summing over $m$ by introducing an expansion
  parameter $x$, we obtain
\beqn
  0 &=& \int \cdots \int \prod_{i} d \lambda_{i}
\left( \sum_{j} \frac{d}{d\lambda_{j}} \frac{1}{x-
  \lambda_{j}} \right) \exp
  \left(- S_{ {\rm eff}}^{(B)} \right)  \;\;, \;\; \nonumber \\
S_{ {\rm eff}}^{(B)} &\equiv&
  - 2 \sum_{i < j}  \log (\lambda_{i} -\lambda_{j})
   + \frac{1}{g_{s}}
  \sum_{i}  W \left( \lambda_{i}  \right) \;\;. \nonumber
\eeqn
  Carrying out the differentiations and straightforward algebras,
   we obtain the  finite $\check{N}$
  loop equation for the resolvent operator
   $\omega_{\check{N}}^{(B)} (x)
\equiv - g_{s} {\displaystyle \sum_{i=1}^{\check{N}}}
\frac{1}{x-\lambda_{i}}$ :
\beqn
0=
 \langle \langle \omega_{\check{N}}^{(B)}(x)^{2}
  \rangle \rangle
  + \langle \langle  W^{\prime}(x)  \omega_{\check{N}}^{(B)} (x)
   \rangle \rangle
  + \langle \langle  \frac{S}{\check{N}}   \sum_{i}
  \frac{W^{\prime}(x) -W^{\prime}(\lambda_{i})}{x-\lambda_{i}}
  \rangle \rangle  \;\;. \nonumber
\eeqn
  Here
$\langle \langle  \cdots  \rangle \rangle$
  denotes the averaging  with respect to  the partition function.
 The absence of the
$g_{s} \langle \langle  \omega_{\check{N}}^{(B)\prime}(x)
  \rangle \rangle$ term is related to $c=1$ nature of the
 Virasoro constraints.
  In the planar limit in which the correlators of the singlet 
operators factorize, we obtain
$ \omega^{(B)}(x) \equiv {\displaystyle \lim_{\check{N} \rightarrow \infty} }
  \omega_{\check{N}}^{(B)} (x)$:
\beqn
\label{planarloop}
    \omega^{(B)}(x)^{2} +
   W^{\prime}(x) \omega^{(B)}(x)  + 
 \lim_{\check{N} \rightarrow \infty}
   \langle \langle  \frac{S}{\check{N}}
 \sum_{i}
  \frac{W^{\prime}(x) -W^{\prime}(\lambda_{i})}{x-\lambda_{i}}
  \rangle \rangle  =0  \;\;. \nonumber
\eeqn

We now apply the above procedure to our model given by
eq. (\ref{ZIK}).  We find that
  the appropriate realization of the Virasoro
   operators
  which involve the grassmann eigenvalues as well is
\beqn
   \ell_{m} &\equiv&  - \sum_{i}
\frac{d}{d\lambda_{i}} \lambda_{i}^{m+1}
   +   \gamma (m+1)
  \sum_{i}  \left( \frac{d}{d \theta_{i}}
  \theta_{i} +
    \frac{d}{d \bar{\theta}_{i}} \bar{\theta}_{i} \right)
    \lambda_{i}^{m}  \;,  \;\; m \geq -1 \;\;  \nonumber  \\
  \hat{\ell} (x) &=& - \sum_{j} \frac{d}{d\lambda_{j}}
  \frac{1}{x- \lambda_{j}}
   + \gamma \sum_{j} \left( \frac{d}{d \theta_{j}}
  \theta_{j} +    \frac{d}{d \bar{\theta}_{j}}
  \bar{\theta}_{j} \right)
  \frac{1}{(x-\lambda_{j})^{2}} \;\;. \label{hatell}
\eeqn
  This form can be motivated  by  realizing the ${\cal N}=2$
   superconformal algebra in terms of
  (super-)differential operators.
  The coefficient $\gamma$ is then determined:
\beq
   \gamma = \frac{1}{2} \;\;.
\eeq

  Inserting $- \hat{\ell}(x)$, which is a total derivative,
  in  $Z_{\check{N}}$, carrying out  algebras,
  and equating the resulting expression to zero, we obtain
\beqn
\label{Vircal}
   &0& = (1- 2\gamma)  \langle \langle
   \sum_{j}
  \frac{1}{(x-\lambda_{j})^{2}}   \rangle \rangle \;\; \nonumber  \\
  &-&\frac{1}{g_{s}}
  \langle \langle
\sum_{j}
  \left( \frac{1}{x- \lambda_{j}} \frac{d}{d\lambda_{j}}
   + \gamma  \frac{1}{(x-\lambda_{j})^{2}}
  \left(   \theta_{j} \frac{d}{d \theta_{j}}
  +   \bar{\theta}_{j}  \frac{d}{d \bar{\theta}_{j}}
  \right)
  \right)
  W(\lambda_{j}+ \theta_{j} \bar{\theta}_{j}) \rangle \rangle \;\;  \\
   &+& 
  2  \langle \langle
\sum_{j}
  \left( \frac{1}{x- \lambda_{j}} \frac{d}{d\lambda_{j}}
   + \gamma   \frac{1}{(x-\lambda_{j})^{2}}
  \left(   \theta_{j} \frac{d}{d \theta_{j}}
   +  \bar{\theta}_{j}   \frac{d}{d \bar{\theta}_{j}}
  \right)    \right)
    \sum_{i < k} \log \left( \lambda_{i} -\lambda_{k} +
  \theta_{i}\bar{\theta}_{i} -\theta_{k}\bar{\theta}_{k}  \right)
   \rangle \rangle     \nonumber  \;\;.
\eeqn
   The first line of this equation vanishes with our choice
$\gamma =\frac{1}{2}$.  The second line with this choice is
\beqn
\label{2ndline}
    - \frac{1}{g_{s}} \langle \langle
  \sum_{j}
\frac{W^{\prime}\left(\lambda_{j} +
\theta_{j}\bar{\theta}_{j} \right)}{x- \lambda_{j} -
\theta_{j}\bar{\theta}_{j} }
   \rangle \rangle  \;\;. \nonumber
\eeqn
  The calculation of the third line is a little less trivial to see.
After antisymmetrization and isolation of purely bosonic  term,
   we obtain
\beqn
    \langle \langle
  \sum_{j \neq i} \frac{1}{(x- \lambda_{j})(x- \lambda_{i})}
   \rangle \rangle
  -  \langle \langle
  \sum_{j \neq i}
  \frac{\theta_{j}\bar{\theta}_{j} -\theta_{i}\bar{\theta}_{i}}
{(x- \lambda_{j})(x- \lambda_{i})
( \lambda_{j}- \lambda_{i}
+ \theta_{j}\bar{\theta}_{j} -\theta_{i}\bar{\theta}_{i})}
   \rangle \rangle  \;\;  \nonumber  \\
  + 2 \gamma
\langle \langle   \sum_{j \neq i}
   \left( \frac{\theta_{j}\bar{\theta}_{j} }{(x- \lambda_{j})^{2} }
-  \frac{\theta_{i}\bar{\theta}_{i}}
{(x- \lambda_{i})^{2} } \right)
  \frac{1}
{ \lambda_{j}- \lambda_{i}
+ \theta_{j}\bar{\theta}_{j} -\theta_{i}\bar{\theta}_{i} }
  \rangle \rangle   \;\;. \nonumber
\eeqn
  The cancellation of the $\frac{1}{x^{2}}$ contribution
  from the second and the third terms of this equation
  in the asymptotic expansion
   again requires $\gamma = \frac{1}{2}$.  With this choice, the
  third line of  eq.(\ref{Vircal})  is
\beq
   \langle \langle
  \sum_{i \neq j} \frac{1}{(x- \lambda_{j}-
   \theta_{j}\bar{\theta}_{j})
  (x- \lambda_{i} -\theta_{i}\bar{\theta}_{i})}
   \rangle \rangle  \;\;.
\eeq
  Introducing the resolvent operator
$\omega_{\check{N}}(x) \equiv
- g_{s} {\displaystyle \sum_{i=1}^{\check{N}} }
\frac{1}{ x-\lambda_{i}-\theta_{i}\bar{\theta}_{i}}$,
  we obtain
\beqn
\label{loopeq}
  0 &=&
  \langle \langle \omega_{\check{N}}(x)^{2}
   \rangle \rangle
   -g_{s} \langle \langle \omega_{\check{N}}^{{\prime}}(x)
   \rangle \rangle
+ \langle \langle  W^{\prime}(x) \omega_{\check{N}} (x)
   \rangle \rangle
  + \langle \langle \frac{S}{\check{N}}   \sum_{i}
  \frac{W^{\prime}(x) -W^{\prime}
(\lambda_{i} +\theta_{i}\bar{\theta}_{i})}
{x-\lambda_{i}-\theta_{i}\bar{\theta}_{i}}
  \rangle \rangle  \;\;\nonumber \\
  &\equiv&
    \langle \langle {\cal L}_{1}(x)  \rangle \rangle  \;\;.
\eeqn
  The fourth term is a polynomial in $x$. In the planar limit
$ \omega(x) \equiv  \lim \omega_{\check{N}}(x)$,
this equation  reduces to
\beq
   \omega(x)^{2} +
   W^{\prime}(x) \omega(x)  + \lim
  \langle \langle
  \frac{S}{\check{N}}  \sum_{i}
  \frac{W^{\prime}(x) -W^{\prime}
(\lambda_{i} +\theta_{i}\bar{\theta}_{i})}
{x-\lambda_{i}-\theta_{i}\bar{\theta}_{i}}   \rangle \rangle
   =0  \;\;.
\eeq

   So far we have managed to obtain a single loop equation
   for the resolvent.
The observables of our model are, on the other hand,
\beqn
  \sum_{i}\lambda_{i}^{\ell}, \;
  \sum_{i}\lambda_{i}^{\ell} \theta_{i}, \;
\sum_{i}\lambda_{i}^{\ell} \bar{\theta}_{i}, \;
\sum_{i}\lambda_{i}^{\ell}  \bar{\theta}_{i} \theta_{i}
  \;\;, \; \ell \geq 0 \;\;. \nonumber
\eeqn
With the single expansion parameter $x$, we are not able to generate
  all of  these which are independent. In order to overcome
  this point,  we insert
\beqn
    \frac{1}{2}
\left( 1+ \eta \sum_{i} \frac{d}{d \theta_{i}}
           +  \sum_{i}\frac{d}{d \bar{\theta}_{i} } \bar{\eta}
   \right)^{2}, \nonumber
\eeqn
following the insertion of
  the Virasoro operators in eq.(\ref{Vircal})
   and carry out the remaining procedure. The grassmann expansion
  parameters $\eta, \bar{\eta}$ have now been introduced.
   Equivalently, we can make a fermionic shift of
   the grassmann eigenvalues in eq.(\ref{loopeq}):
\beqn
\theta_{i}  \rightarrow  \theta_{i}+  \eta \;,\;\;
\bar{\theta}_{i}  \rightarrow   \bar{\theta}_{i} + \bar{\eta}\;\;.
 \nonumber
\eeqn
  Both the resolvent operator
   $\omega_{\check{N}} (x)$ and
\beq
S_{ {\rm eff}} \equiv
\frac{1}{g_{s}}
  \sum_{i}  W (\lambda_{i} + \theta_{i}\bar{\theta}_{i})
  -2
  \sum_{i < j}  \log ( \lambda_{i} -\lambda_{j} +
  \theta_{i}\bar{\theta}_{i} -\theta_{j}\bar{\theta}_{j} )
    \nonumber
\eeq
undergo  this shift while
$  d (\theta_{i}+  \eta) = d \theta_{i}   \;,\;\;
d(\bar{\theta}_{i} + \bar{\eta}) = d\bar{\theta}_{i}$.
  In either way, we obtain another set of Schwinger-Dyson
   equations as  equations multiplying $\eta, \bar{\eta},
\eta\bar{\eta}$;
\beqn
\label{loopeq23}
  0 &=&   \langle \langle  {\cal L}_{2}(x)
  +  
  {\cal S}_{\theta}^{{\rm sub}} 
  {\cal L}_{1}(x)    \rangle \rangle
 \;\;, \nonumber
\\
  0 &=&    \langle \langle  {\cal L}_{\bar{2}}(x)
    +  
  {\cal S}_{\bar{\theta}}^{{\rm sub}} 
   {\cal L}_{1}(x)   \rangle \rangle 
   \;\;,   \\
  0 &=&   \langle \langle 
  {\cal L}_{3}(x) + 
 \left(
 {\cal S}_{ \theta \bar{\theta}}^{{\rm sub}}
 + 
{\cal S}_{\theta}^{{\rm sub}}{\cal S}_{\bar{\theta}}^{{\rm sub}}  
-  
\langle \langle {\cal S}_{\theta}^{{\rm sub}} 
{\cal S}_{\bar{\theta}}^{{\rm sub}} \rangle \rangle
\right) 
   {\cal L}_{1}(x) 
+
  {\cal L}_{\bar{2}}(x)  {\cal S}_{\bar{\theta}}^{{\rm sub}}
+ 
 {\cal S}_{\theta}^{{\rm sub}}{\cal L}_{2}(x)  \rangle \rangle
 \;\;, \nonumber
\eeqn
where
\beqn
{\cal L}_{2}(x) &\equiv&
2  \psi_{\check{N}}(x)  \omega_{\check{N}} (x)
   -g_{s}  \psi_{\check{N}}^{\prime}(x)
   +  W^{\prime}(x) \psi_{ \check{N}} (x)
 +   \rho_{ \check{N}}(x)  \nonumber
~, \\
{\cal L}_{\bar{2}}(x) &\equiv&
2  \bar{\psi}_{\check{N}}(x)  \omega_{\check{N}} (x)
   -g_{s}  \bar{\psi}_{\check{N}}^{\prime}(x)
  +  W^{\prime}(x) \bar{\psi}_{\check{N}} (x)
  +   \bar\rho_{ \check{N}}(x) 
~, \\
{\cal L}_{3}(x) &\equiv&
2  t_{\check{N}}(x)  \omega_{\check{N}} (x)
+  2  \bar{\psi}_{\check{N}}(x) \psi_{\check{N}}(x)
   -g_{s}  t_{ \check{N}}^{\prime}(x)
+  W^{\prime}(x) t_{ \check{N}} (x)
  +  c_{ \check{N}}(x) \;\;, \nonumber
\eeqn
and
\beqn
\label{rhoc}
\rho_{ \check{N}}(x) &=&  \frac{S}{\check{N}}
\sum_{i} \theta_{i}
 \frac{ 
W^{\prime (2)}_{{\rm sub}}
(x;\lambda_{i} +\theta_{i}\bar{\theta}_{i})}
{\left(x-\lambda_{i}-\theta_{i}\bar{\theta}_{i} \right)^{2}}~,
\;\;
\bar\rho_{ \check{N}}(x) = \frac{S}{\check{N}}
\sum_{i}  \bar{\theta}_{i}
  \frac{
W^{\prime (2)}_{{\rm sub}}
(x;\lambda_{i} +\theta_{i}\bar{\theta}_{i})}
{\left(x-\lambda_{i}-\theta_{i}\bar{\theta}_{i} \right)^{2}}~,
  \nonumber \\
c_{ \check{N}}(x) &=&   \frac{S}{\check{N}}
\sum_{i}
  \frac{
W^{\prime (2)}_{{\rm sub}}
(x;\lambda_{i} +\theta_{i}\bar{\theta}_{i}) }
{\left(x-\lambda_{i}-\theta_{i}\bar{\theta}_{i} \right)^{2}}
  + \frac{S}{\check{N}} \sum_{i}  2\bar{\theta}_{i} \theta_{i}
  \frac{
W^{\prime (3)}_{{\rm sub}}
(x;\lambda_{i} +\theta_{i}\bar{\theta}_{i})
 }
{\left(x-\lambda_{i}-\theta_{i}\bar{\theta}_{i} \right)^{3}} \;\;.
\eeqn
 We have introduced
\beq
W^{\prime (m+1)}_{{\rm sub}}
(x;\lambda_{i} +\theta_{i}\bar{\theta}_{i}) \equiv
W^{\prime}(x) -
  \sum_{\ell=0}^{m}  \frac{(x-\lambda_{i} -\theta_{i}\bar{\theta}_{i})^{\ell} }{\ell !} 
W^{(\ell+1)}
(\lambda_{i} +\theta_{i}\bar{\theta}_{i})\;\;.  \nonumber
\eeq 
We see that the quantities seen in eq.(\ref{rhoc}) 
are polynomials in $x$.
 Another bosonic resolvent and
  two fermionic resolvents are defined respectively by
\beqn
t_{\check{N}}(x) &\equiv&
- g_{s} \sum_{i}
  \left( \frac{1}{ \left(x-\lambda_{i}-\theta_{i}\bar{\theta}_{i}
  \right)^{2}}  + \frac{2  \bar{\theta}_{i} \theta_{i} }
{ \left(x-\lambda_{i}-  \theta_{i}\bar{\theta}_{i}
  \right)^{3}} \right),  \nonumber \\
\psi_{\check{N}}(x) &\equiv&
- g_{s} \sum_{i}
\frac{ \theta_{i} }{ \left(x-\lambda_{i}
-\theta_{i}\bar{\theta}_{i}\right)^{2}}, \;\;
\bar{\psi}_{\check{N}}(x) \equiv
- g_{s} \sum_{i}
\frac{\bar{\theta_{i}}}
{ (x-\lambda_{i}-\theta_{i}\bar{\theta}_{i})^{2}} \;. \nonumber
\eeqn

   In the planar limit, the terms involving
 the singlet operators coming from the
  variation of the action  which we have  denoted by
\beqn
  {\cal S}_{\theta} \equiv - \sum_{i} \frac{\partial S_{\rm eff}}
   {\partial \theta_{i}}\;, \;\;
  {\cal S}_{\bar{\theta}} \equiv
 - \sum_{i} \frac{\partial S_{\rm eff}}
   {\partial \bar{\theta_{i}} }\;, \;\; 
  {\cal S}_{ \theta \bar{\theta}}  \equiv -
\sum_{i,j} 
  \frac{  \partial^{2} S_{\rm eff}}
{ \partial \theta_{j} \partial \bar{\theta_{i}} }
 \;,  \;\;
 {\cal S}_{\bullet}^{{\rm sub}} \equiv    {\cal S}_{\bullet}
  - \langle \langle {\cal S}_{\bullet} \rangle \rangle \;\;
\eeqn 
  factorize and become ignorable.
   We obtain
\beqn
\label{planarloopeq23}
  &&  2  \psi(x)  \omega(x)
  +  W^{\prime}(x) \psi(x)
  +  \lim \langle \langle  \rho_{ \check{N}}(x)
   \rangle \rangle  =0\;, \;\;  \nonumber \\
   &&
2  \bar{\psi}(x)  \omega (x)
+   W^{\prime}(x) \bar{\psi} (x)
  +  \lim  \langle \langle \bar\rho_{ \check{N}}(x)
   \rangle \rangle =0 \;,
   \;\;  \\
  &&
2  t(x)  \omega (x)
+  2  \bar{\psi}(x) \psi(x)
+  W^{\prime}(x) t(x)
  +  \lim \langle \langle   c_{ \check{N}}(x)
\rangle \rangle   =0 \;.  \nonumber
\eeqn
 As is announced, the system of  three planar loop obtained 
 in eq.(\ref{planarloopeq23})
  takes  the same form as that of the equations of
  Konishi anomaly in \cite{CDSW}.

  So far, we have dealt with  the case in which the underlying
  gauge group is $U(N)$; this has led us to  the supereigenvalue
  generalization of  the hermitian one-matrix model.
 It is easy to  generalize to 
  the cases  in which  the underlying gauge group is
 $SO(2N)/Sp(2N),SO(2N+1)$.  
(See the recent discussion \cite{ACHKR}).
 The measure factor is changed into
\beqn
     \prod_{i < j}^{\check{N}} ( \lambda_{i} -\lambda_{j} +
  \theta_{i}\bar{\theta}_{i} -\theta_{j}\bar{\theta}_{j} )^{2}
  ( \lambda_{i} +\lambda_{j} +
  \theta_{i}\bar{\theta}_{i} +\theta_{j}\bar{\theta}_{j} )^{2}
  \;, &\;\;& {\rm for} \; SO(2{\check{N}})     \nonumber \\
     \prod_{i < j}^{\check{N}} ( \lambda_{i} -\lambda_{j} +
  \theta_{i}\bar{\theta}_{i} -\theta_{j}\bar{\theta}_{j} )^{2}
  ( \lambda_{i} +\lambda_{j} +
  \theta_{i}\bar{\theta}_{i} +\theta_{j}\bar{\theta}_{j} )^{2}
   \prod_{i}^{\check{N}} ( \lambda_{i}  +
  \theta_{i}\bar{\theta}_{i} )^{2}
 \;, &\;\;& {\rm for}\; Sp(2 {\check{N}}), SO(2 {\check{N}} +1)
  \;\;.    \nonumber
\eeqn
 These cases can be  handled by  changing  the set which integer
  labels $i,j$ have belonged to into the new ones:
\beqn
  i,j \in \{ 1, 2, \cdots, \check{N} \}
  &\Rightarrow&     i,j \in \{ -\check{N},  \cdots, -2, -1, 1, 2,
  \cdots, \check{N} \} \;, \;\; {\rm for} \; SO(2 {\check{N}})    \nonumber \\
  i,j \in \{ 1, 2, \cdots, \check{N} \}
  &\Rightarrow &    i,j \in \{ -\check{N},  \cdots, -2, -1, 0, 1, 2,
  \cdots, \check{N} \} \;, \;\; {\rm for}\; Sp(2 {\check{N}}),
 SO(2 {\check{N}})    \nonumber 
\eeqn
 and by demanding
\beqn
  \lambda_{-i} = - \lambda_{i} \;, \;\; \lambda_{0} = 0 \;\;.  \nonumber
\eeqn
 The summations and the products in the previous equations
  are now taken  with respect the integers
  belonging to these new sets. The potential effectively becomes
a polynomial consisting only of even powers. The remainder of the
  discussions are the same.

 Finally, let us  discuss the supereigenvalue model
 which implements ${\cal N}=2$ super Virasoro constraints.
 The action is given by the ${\cal N}=2$ superfield (\ref{bfV}).
 In the partition function  (\ref{Z}), the measure factor is 
replaced by
\beq
\prod_{i < j} ( \lambda_{i} -\lambda_{j} -
 (\theta_{i}\bar{\theta}_{j} + \bar{\theta}_{i} \theta_{j}) )^{2}
  \;\;,
\eeq
 where  each factor is the super-distance invariant under
  the supertranslation generated by
the supercovariant derivatives.
 The (super)-differential operators
to be inserted in the partition function are
\beqn
\hat j(x) &=& \sum_j \left(  \frac{\p}{\p \theta_j} \theta_j - 
\frac{\p}{\p \Bar\theta_j} \Bar\theta_j \right) \frac{1}{x- 
\lambda_j}~,  \nonumber \\
\hat g(x) &=&  \sum_j \left(  \frac{\p}{\p \theta_j}  -  \Bar 
\theta_j  \frac{\p}{\p \lambda_j} \right) \frac{1}{x- \lambda_j - 
\theta_j \Bar\theta_j }~,   \\
\hat {\Bar g}(x) &=& \sum_j \left(  \frac{\p}{\p \Bar\theta_j} - 
\theta_j  \frac{\p}{\p \lambda_j} \right) \frac{1}{x- \lambda_j - 
\Bar\theta_j \theta_j}~, \nonumber
\eeqn
 and $\hat \ell (x)$ (the case $\gamma = \frac{1}{2}$) 
defined in eq. (\ref{hatell}).
 They can be organized into a current superfield.
 The same is true for the resolvents of this model:
\beq
\Omega(x, \eta, \Bar\eta) \equiv  -g_{s} 
\sum_j \frac{(\eta - \theta_j)(\Bar\eta - 
\Bar\theta_j)}{x-\lambda_j}~.
\eeq
 Introducing  the supercovariant derivatives
$ D \equiv  \frac{\p}{\p \eta} +  \Bar\eta \frac{\p}{\p x}
\; , \;\;
\Bar D \equiv  \frac{\p}{\p \Bar\eta} +  \eta \frac{\p}{\p x} $,
 we have been able to derive a manifestly ${\cal N}=2$ 
supersymmetric
  loop equation. Space only permits us to present its final form:
\beq
0 =  \langle\langle D\Omega \Bar D\Omega \rangle\rangle 
  - g_{s} 
 \langle\langle \Omega^{\prime} \rangle \rangle 
 + \langle\langle \left( D{\bf V} \Bar D \Omega 
+ \Bar D{\bf V} D \Omega \right) \rangle\rangle
+  \langle\langle {\cal {\bf F}} \rangle\rangle~.
\eeq
 Here  ${\cal {\bf F}}(x, \eta, \Bar\eta)$ is a polynomial
in $x$ given explicitly in terms of the components of ${\bf V}$.
Factorization of the two-point functions
in the large  $\check{N}$ limit ($g_{s} \rightarrow 0$)
 gives us an ${\cal N}=2$ planar loop equation.

We plan to give  full details of the contents of this letter
  together with other issues in a future publication.
 We thank  Alyosha Morozov  for helpful discussion
  on this subject.




\end{document}